\documentclass{article}

\usepackage[nonatbib]{arxiv}

\usepackage[pdftex]{graphicx}
\graphicspath{{../pdf/}{../jpeg/}}
\DeclareGraphicsExtensions{.pdf,.jpeg,.png}
\usepackage{amsmath}
\usepackage{amsfonts}
\usepackage{array}
\usepackage{url}
\usepackage{bm}
\usepackage{algorithmic}
\usepackage{arydshln}  
\usepackage{multirow}
\usepackage{hyperref}

\makeatletter
\def\endthebibliography{%
	\def\@noitemerr{\@latex@warning{Empty `thebibliography' environment}}%
	\endlist
}
\makeatother
\def\I{\mathrm{i}}
\newcommand{\norm}[1]{\left\|#1\right\|}

\begin{document}

\title{Estimating the material parameters of an inhomogeneous poroelastic plate from ultrasonic measurements in water}

\author{Matti~Niskanen$^{a,b,}$\thanks{Corresponding author.
		email: \emph{matti.niskanen@uef.fi}}\hspace{1ex},
	Aroune~Duclos$^{b}$,
	Olivier~Dazel$^{b}$,
	Jean-Philippe~Groby$^{b}$,
	Jari~Kaipio$^{c}$,
	Timo~L\"{a}hivaara$^{a}$}

\date{$^{a}$ Department of Applied Physics, University of Eastern Finland, Kuopio, Finland\\%
	$^{b}$ Laboratoire d'Acoustique de l'Universit\'{e} du Mans, LAUM - UMR CNRS 6613, Le Mans, France\\%
	$^{c}$ Department of Mathematics, University of Auckland, Auckland, New Zealand\\[2ex]%
	\today}

\maketitle

\begin{abstract}
The estimation of poroelastic material parameters based on ultrasound measurements is considered.
The acoustical characterisation of poroelastic materials based on various measurements is typically carried out by minimising a cost functional of model residuals, such as the least squares functional.
With a limited number of unknown parameters, least squares type approaches can provide both reliable
parameter and error estimates.
With an increasing number of parameters, both the least squares parameter estimates and, in particular, the error estimates often become unreliable.
In this paper, the estimation of the material parameters of an inhomogeneous poroelastic (Biot) plate in the
Bayesian framework for inverse problems is considered.
Reflection and transmission measurements are performed and 11 poroelastic parameters, as well as 4 measurement setup-related nuisance parameters, are estimated.
A Markov chain Monte Carlo algorithm is employed for the computational inference to assess the actual uncertainty of the estimated parameters.
The results suggest that the proposed approach for poroelastic material characterisation can reveal the heterogeneities in the object, and yield reliable parameter and uncertainty estimates.
\end{abstract}

\vspace{1cm}
\section{Introduction}

Methods for characterising poroelastic media are needed in a wide range of applications, such as studying living bone tissue \cite{cowin1999bone}, characterising seabed \cite{chotiros2017acoustics}, geophysical exploration \cite{slatt2013stratigraphic}, design of materials for noise treatment \cite{allard2009propagation,cox2016acoustic}, and industrial filtration \cite{espedal2007filtration}.
The inverse problem of estimating the physical parameters of such media from acoustic measurements has received a lot of interest \cite{sebaa2006ultrasonic,jocker2009ultrasonic,buchanan2011recovery,ogam2011non,chazot2012acoustical,verdiere2017inverse,chen2018biot}, in part because the acoustic measurements are non-destructive, and the measurement set up is simple compared to many direct measurements of the material properties.
A concern that remains in using inverse methods for the characterisation is that currently the reproducibility seems to be poor even when estimating only the properties of rigid frame porous media, as shown in a recent study \cite{pompoli2017reproducible}, or only the elastic properties, see \cite{bonfiglio2018reproducible}.
Furthermore, usually a larger number of unknowns leads to larger uncertainty in the estimates.

Early approaches to the inverse problem were deterministic, i.e. concentrated on finding the best fit parameters by minimising a cost function, or by finding the analytical inverse mapping (if it exists) that connects data to parameters.
With linear problems, and problems that can be reasonably linearised, deterministic approaches can produce reliable parameter and uncertainty estimates, and often with a relatively low computational cost.
On the other hand, deterministic approaches do not provide a general way of treating errors in nonlinear problems or specifying prior information.
Note that all characterisation methods in the two reproducibility studies \cite{pompoli2017reproducible,bonfiglio2018reproducible} are deterministic.
It has also been proposed \cite{chazot2012acoustical} to treat the inverse problem in the statistical (Bayesian) framework, where instead of trying to find a single value for the parameters, we are looking for their \emph{posterior probability distribution} (ppd) \cite{kaipio2006statistical}.
The ppd is constructed based on the measured data by assessing the uncertainty in the measurements, and by incorporating possible prior knowledge on the parameters.
We can then calculate the most likely parameter values, as well as their credible interval estimates.
Despite the interest in characterising poroelastic materials, only a few studies that consider Bayesian inversion exist so far.
Bayesian inversion has been, however, widely used in characterisation of media that can be modelled as an equivalent fluid, such as rigid frame porous materials \cite{niskanen2017deterministic,roncen2018bayesian}, and seabed \cite{dettmer2007full,dettmer2010trans,dettmer2012trans,dettmer2013transdimensional,holland2013situ}.

Bayesian inversion was used by Chazot \emph{et al.} \cite{chazot2012acoustical} to study the characterisation of highly porous foam and fibre materials using measurements done in an impedance tube.
To see the elasticity effects in the impedance tube, the materials need to be relatively soft, which means that the method is suitable for mostly foams and fibrous materials.
Bonomo \emph{et al.} \cite{bonomo2018comparison} used the Bayesian approach to compare three poroelastic models for sandy sediments and infer their parameters, using either compressional or shear wave speed data acquired from water saturated fine-grained silica sand.
It was found that while the models could be used to explain most of the measurements, the parameters related to the elastic behaviour of the solid were mostly unidentifiable from the data.
It was concluded that the Biot model is not directly suitable for modelling sandy materials since the unconsolidated granular medium does not form an effective solid frame.
Niskanen \emph{et al.} \cite{niskanen2019characterising} considered Bayesian inversion numerically in the case of poroelastic materials that have a solid porous frame such as limestone, trabecular bone, or porous ceramic.
Simulation data for the inversion were acoustic reflection and transmission coefficients measured at low ultrasound frequencies.
The study also considered errors related to the measurement setup, and showed that reliable parameter estimates can be found to all parameters in the Biot model, even in the presence of relatively high levels of noise.
However, the study mainly considered additive white measurement noise whose characteristics are known, whereas with real measurements the noise term includes modelling errors that are often known approximately at best.

In this paper, we apply the inversion method in \cite{niskanen2019characterising} to characterise a porous ceramic plate using real measurements made in a water tank.
Using focused ultrasound transducers, we measure the normal incidence acoustic reflection and transmission coefficients of the plate at more than 200 locations.
These measurements are linked to the physical parameters using the Biot-Johnson model \cite{johnson1987theory}, and a Global Matrix Method solution, which assumes plane wave propagation.
Then, we use Bayes' formula to construct the ppd, and compute the parameter estimates and their uncertainties with a Markov chain Monte Carlo sampling algorithm.
We find that the plane wave model can produce model predictions that fit the actual measurements, and like in the numerical case \cite{niskanen2019characterising}, data carry information on the variables in the sense that posterior variance is reduced compared to the prior variance.
We also find that the proposed method can find the inhomogeneities in the specimen, for each parameter separately.

The rest of the paper is organised as follows.
First we describe the measurement set-up in section \ref{sec:experiments}.
Next we discuss the modelling of the poroelastic wave propagation in section \ref{sec:forwardproblem}.
Then, in section \ref{sec:inverse_problem} we formulate the inverse problem and consider some uncertainties in the measurements.
Results are presented and discussed in section \ref{sec:results}, followed by a conclusion in section \ref{sec:conclusion}.

\section{Experiment and data processing}
\label{sec:experiments}

We are interested in testing the inversion method on a porous material that is stiff, so that wave propagation in air cannot move its frame, and the material needs to be submerged in water to excite the solid phase.
Porous ceramics usually fit such description and we therefore choose the porous ceramic QF-20, which is made of glass bonded silica and manufactured by Filtros Ltd.
An approximately rectangular, parallel-faced, plate cut of the material is submerged in a water tank, and proper precaution is taken to ensure the pores are saturated with water.

Data needed for the inversion are the plate's acoustic reflection and transmission coefficients, which in the frequency domain we denote by $\bm{R}$ and $\bm{T}$, respectively.
The basic idea of the measurement is that a known wave pulse is sent towards the object, and the waves reflected back and transmitted through the material are recorded.
These can be measured with ultrasound transducers in two measurement configurations, one for reflection and another for transmission.
By comparing the incident and recorded waves we can estimate the desired acoustic responses.
A schematic of the set up is shown in Fig.~\ref{fig:watertank_render}.

To keep the forward model simple we assume plane wave propagation and homogeneity of the material, as discussed in section~\ref{sec:forwardproblem}.
Therefore, the obvious choice for the experiments is to measure the plane wave reflection and transmission coefficients.
To realise conditions resembling plane waves, we could use very large transducers \cite{castaings2000inversion} or synthetic plane wave techniques \cite{jocker2007minimization}.
However, these methods average the measurements over a large area, and small scale resolution of the object would be lost.
In addition, if the object is inhomogeneous in the macroscopic sense, the forward model would be incompatible with measurements averaged over regions with differing physical properties.
It would therefore be advantageous to measure as small a portion of the plate at a time as possible, and be able to locally approximate the plane wave condition.
Being able to take multiple measurements from different places of the same object allows us not only to assess the homogeneity of the material, but also the validity of the inversion method.
In this work, we use focused transducers for the reason that they have a small footprint on the plate, and that the waveform they send spreads less than the waveform from a flat transducer, approximating plane wave conditions better.

\subsection{The measurement system}

\begin{figure}[t]
	\centering
	\includegraphics[width=0.50\linewidth]{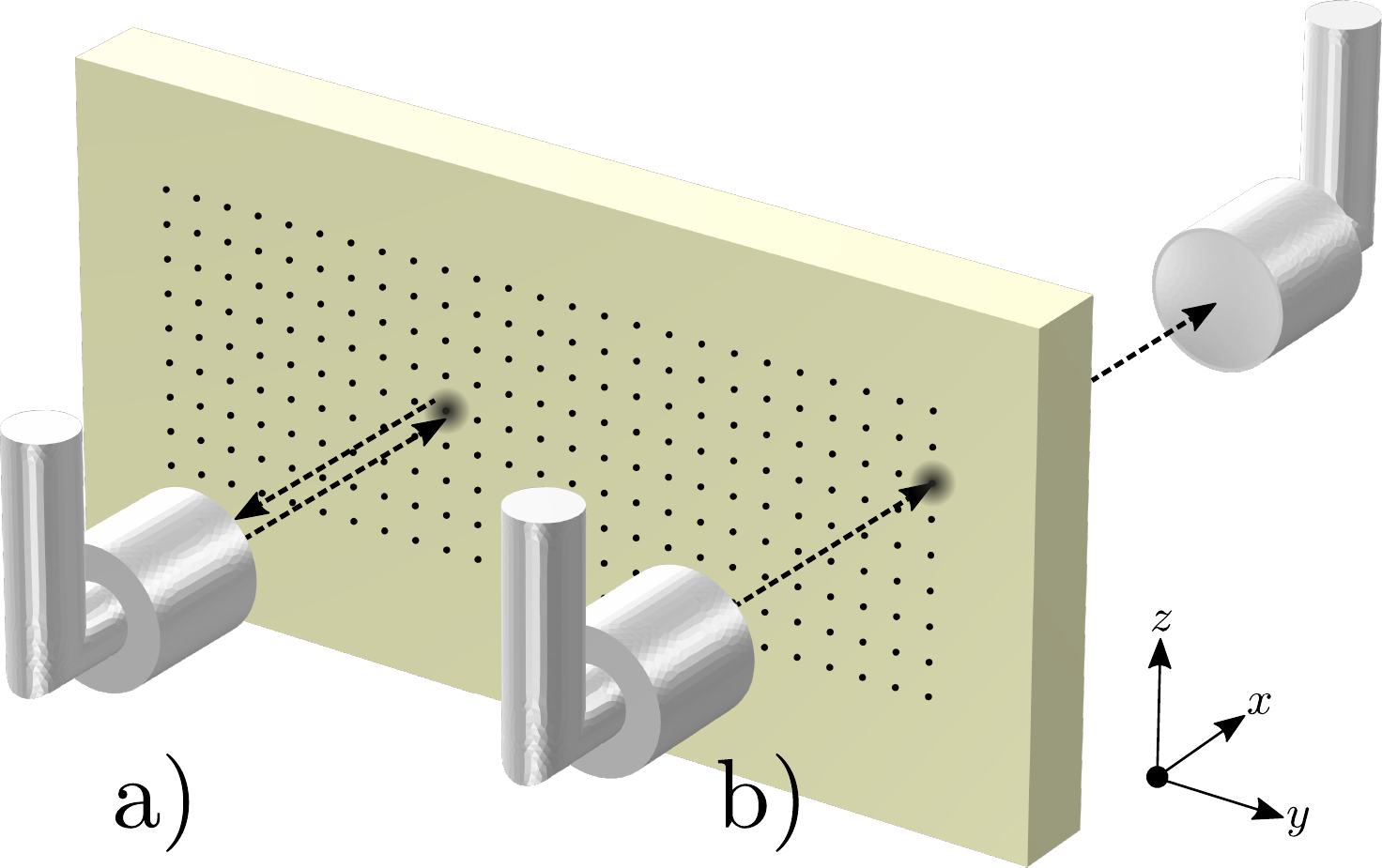}
	\caption{A schematic of the a) reflection, and b) transmission measurement setups. The dots on the plate denote the locations that were measured, and the shading is an indication of the -6 dB beam width.}
	\label{fig:watertank_render}
\end{figure}

We first consider the frequency range of the measurements.
In order to see the effects of the Biot waves (i.e. waves propagating both in the frame and fluid), the frequency needs to be below the regime where scattering effects dominate.
For QF-20, based on conducting tests on the model fit and considering the validity the long-wavelength condition, we approximate the scattering regime to start from 400 kHz onwards.
The front and back interfaces of the plate are smooth down to the pore scale so that we do not have separate scattering from roughness of the interfaces.
The low frequency limit is in practice defined by the frequency response of the transducer, since the Biot model is valid down to audible frequencies.
To produce low-frequency ultrasound we use a point-focused piezoelectric transducer (model Panametrics V389) as both the transmitter and the receiver.
This broadband transducer has a diameter of 38 mm, focal length of 96 mm, and the central frequency at 500 kHz, which in water corresponds to the transducer diameter per wavelength ($D/\lambda$) of 12.
A practical lower frequency limit for the transducer is 200 kHz, where it is still directive ($D/\lambda \approx 5$) and the pulse-echo signal is attenuated by 14 dB compared to the peak.
The -6~dB beam diameter of a pulse-echo signal at the transducer's focal point is 10 mm at 400 kHz \cite{fowler2012important}.

The transducers are rigidly connected to a computer-controlled $XYZ$-positioning system.
The largest face of the measured object is aligned along the $yz$-plane, and the transducers are angled at normal incidence to the same plane (see Fig.~\ref{fig:watertank_render}).
To maximise the signal strength we put each transducer at a distance of 105 mm from the nearest face of the measured object, where the object is in the transducer's focal zone.

As the signal source we use a Sofranel 5072PR pulser/receiver, which can act both in a through-transmission and a pulse-echo mode.
The source works by sending a -360 V spike excitation to the emitting transducer.
The signal from the receiving transducer is then sent through the pulser/receiver, without any analogue filtering, to a Picoscope 5244B 16 bit AD-converter.
The received signals show contributions from two waves with different velocities, which we can attribute to the fast and the slow longitudinal wave predicted by the Biot theory.
This confirms that the measurements carry information on the Biot effects.
Finally, the received signals are digitally averaged over 40 pulses, to reduce the effect of random electrical noise.


The computer positioning system is used to move the transducers and repeat the measurement procedure over a 90$\times$240 mm grid, with 10 mm spacing.
This results in 216 measurement locations, represented by the black dots in Fig.~\ref{fig:watertank_render}.
The spacing between the measurement points is chosen small to achieve a high spatial resolution of the object, but not smaller than the -6 dB beam width to avoid making redundant measurements.

\begin{figure}[t]
	\centering
	\includegraphics[width=0.50\linewidth]{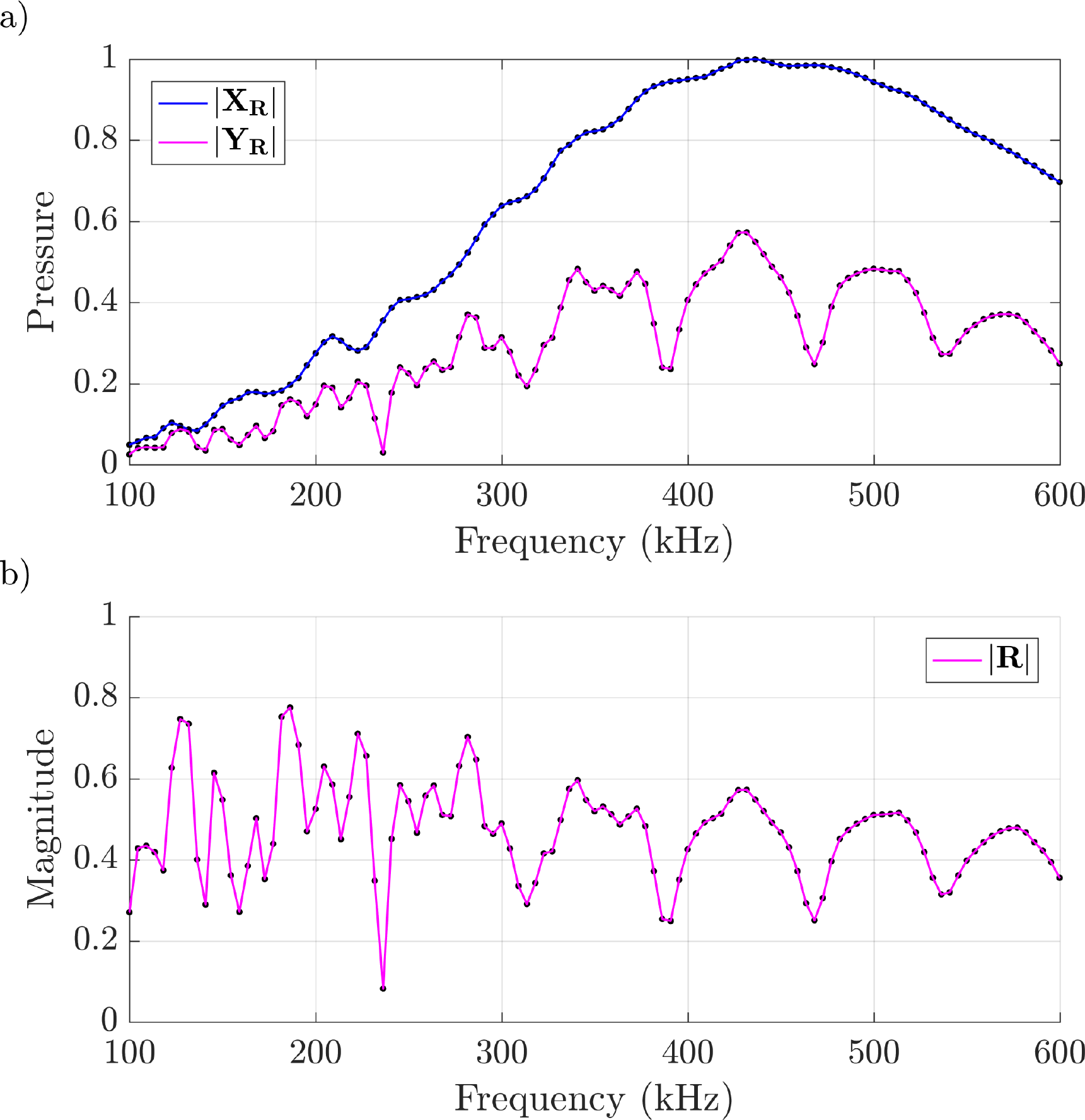}
	\caption{Magnitude of the a) measured incident and reflected fields, b) calculated reflection coefficient, as a function of frequency.}
	\label{fig:Wienerfilter}
\end{figure}

\subsection{Spectral ratio technique}

Let us model a measurement in the reflection configuration, Fig.~\ref{fig:watertank_render}~a), as

\begin{equation}
\bm{y}_R(t) = \bm{x}_R(t)*\bm{h}_R(t) + \bm{n}(t) \:,
\end{equation}

\noindent where $\bm{x}_R(t)$ is the incident wave sent by a transducer towards the plate, $\bm{y}_R(t)$ is a measurement of the wave that is reflected back (including multiple reflections from within the plate), $\bm{h}_R(t)$ is the plate's impulse response, the operator $*$ denotes convolution, $\bm{n}(t)$ is random measurement noise, and $t$ denotes time.
In the reflection configuration we use only one transducer that operates in a pulse-echo mode, so that it both sends and receives the acoustic waves.

The impulse response $\bm{h}_R(t)$ is related to the reflection coefficient of the plate by the Fourier transform $\bm{R}(\omega) = \int_{-\infty}^{\infty}\bm{h}_R(t)e^{-\I\omega t}dt$, where $\omega = 2\pi f$ and $f$ denotes frequency.
In the presence of noise, deconvolution to find $\bm{R}(\omega)$ can be done for example using the Wiener filter \cite{chen1990effective}:

\begin{equation} \label{eq:Wiener_filter_R}
\bm{R}(\omega) = \dfrac{\bm{Y}_R(\omega)\bm{X}_R^*(\omega)}{|\bm{X}_R(\omega)|^2 + q} \:,
\end{equation}

\noindent where $\bm{Y}_R(\omega)$ and $\bm{X}_R(\omega)$ are the Fourier transforms of $\bm{y}_R(t)$ and $\bm{x}_R(t)$, respectively, $^*$ denotes the complex conjugate, and $q$ is the variance of the noise, sometimes also called the noise desensitising factor, that regularises the filter in frequencies where the signal-to-noise ratio is low.

In order to calculate the reflection coefficient \eqref{eq:Wiener_filter_R} accurately, we need to also measure the incident wave transmitted by the source, since the transducer response greatly affects the signal.
If both $\bm{y}_R(t)$ and $\bm{x}_R(t)$ are measured using exactly the same system, the response of the measurement system mostly cancels out.
The incident signal $\bm{x}_R(t)$ can be measured by pointing the transducer upwards to the air-water interface, which in theory gives a total reflection due to the high specific impedance difference between water and air.
The 180 degree phase difference from the water-air reflection needs to be accounted for before further analysis.
An example of the magnitude of the measured $\bm{Y}_R(\omega)$, $\bm{X}_R(\omega)$, and $\bm{R}(\omega)$, calculated using \eqref{eq:Wiener_filter_R}, is shown in Fig.~\ref{fig:Wienerfilter}.

In transmission measurements, Fig.~\ref{fig:watertank_render} b), we use two transducers operating in through-transmission mode, and a reference signal $\bm{x}_T(t)$ is recorded in the same configuration but with the plate removed.
Otherwise the measurement is modelled in the same way as in the reflection case, and the transmission coefficient is obtained as

\begin{equation} \label{eq:Wiener_filter_T}
\tilde{\bm{T}}(\omega) = \dfrac{\bm{Y}_T(\omega)\bm{X}_T^*(\omega)}{|\bm{X}_T(\omega)|^2 + q}e^{-\I k_f L} \:,
\end{equation}

\noindent where $k_f = \omega/c_f$ is the wavenumber in water, $c_f$ speed of sound in water, and $L$ is the thickness of the plate.
The exponential term accounts for the phase difference introduced when the object is removed from the signal path.
Since the term includes the plate's thickness, which we model as one of the unknown parameters and hence do not know prior to the inversion, what we actually measure and use in the inversion is $\bm{T}(\omega) := \tilde{\bm{T}}(\omega)e^{\I k_f L}$.

\section{The forward problem}
\label{sec:forwardproblem}

In this section, let us briefly discuss the forward problem, i.e. how we describe wave motion in poroelastic media and how the theoretical reflection and transmission coefficients are obtained from a set of model parameters.
A more comprehensive treatment of the related equations and solution methods is outside the scope of the current work, and can be found for example in \cite{gautier2011propagation,niskanen2019characterising}.

As is usually done, we model poroelastic media after the Biot theory \cite{biot1956theoryLow,biot1956theoryHigh,biot1962mechanics}, where the material is seen to consist of two interlinked phases, a porous solid frame and a fluid saturating the pores.
Our numerical model consists of three homogeneous and isotropic layers, where a poroelastic layer is sandwiched between two water layers extending to infinity.
Waves are assumed to be propagating normally to the interfaces, making the problem effectively one-dimensional.
When the physical properties of the fluid and poroelastic media are known, the theoretical plane wave transmission and reflection coefficients of the system can be computed by solving the Biot equations.
This can be done, for example, by using the Global Matrix Method \cite{knopoff1964matrix,lowe1995matrix}.
The basic Biot model requires several input parameters: open porosity $\phi$, static viscous permeability $k_0$, geometric tortuosity $\alpha_\infty$, bulk modulus of the solid frame $K_b$, bulk modulus of the solid from which the frame is made of $K_s$, shear modulus of the frame $N$, density of the solid $\rho_s$, bulk modulus of the fluid $K_f$, density of the fluid $\rho_f$, and dynamic viscosity of the fluid $\eta$.
We assume that the properties of the saturating fluid are known, and set $K_f = 2.19$ GPa, $\rho_f = 1000$ kg$\cdot$m$^{-3}$, and $\eta = 1.14\cdot 10^{-3}$ Pa$\cdot$s.

Several models are available to represent attenuation of waves propagating in a poroelastic medium.
The two main types of attenuation are related to viscous losses due to movement of the fluid and to viscous losses in the solid frame.
Attenuation in the fluid can be accounted for by the dynamic tortuosity model of Johnson \emph{et al.} \cite{johnson1987theory}, which introduces another parameter, viscous characteristic length $\Lambda$.
A common way to represent losses in the solid is to give the elastic constants $K_b$, $K_s$, and $N$ a small imaginary part, as was done in \cite{niskanen2019characterising}.
However, an attenuating model with constant real and imaginary parts can be shown to be weakly non-causal \cite{turgut1991investigation}, and in reality the real and imaginary parts of the elastic moduli should be frequency dependent.

In this work, we adopt the Kjartansson model \cite{kjartansson1979constantq}, which satisfies the causality requirement while only having two independent parameters.
Dissipation in the solid can be quantified by the quality factor $Q(\omega)$, which is defined as $Q(\omega) = M_R(\omega)/M_I(\omega)$, the ratio of the real part to the imaginary part of a general elastic modulus $M = M_R + \I M_I$ \cite{bourbie1987acoustics}.
In the Kjartansson model the quality factor is constant over the frequencies, and elastic moduli are represented as

\begin{equation}
M(\omega) = M_0(\I\omega/\omega_0)^{\frac{2}{\pi}\tan^{-1}(1/Q)},
\end{equation}

\noindent where $\omega_0$ is an arbitrary reference frequency, $M_0$ is the value of the elastic modulus at $\omega_0$, and $Q^{-1}$ is the specific attenuation.
Highly attenuating materials have a small $Q$, and vice versa.
In our model, each elastic modulus is now represented by a reference value and a quality factor, i.e. we have $K_{b,0}$ and $Q_{K_b}$, $K_{s,0}$ and $Q_{K_s}$, as well as $N_0$ and $Q_N$.

\section{The inverse problem}
\label{sec:inverse_problem}

The Bayesian approach for the present inverse problem was numerically studied in \cite{niskanen2019characterising}.
In the following, we present an overview of the inversion method, and for the implementation details see \cite{niskanen2019characterising}.
For general references on Bayesian inversion, see \cite{kaipio2006statistical,calvetti2007introduction,tarantola2005inverse}.

The solution of an inverse problem in the Bayesian framework is the ppd, the probability density of the unknown parameters conditioned on the measured data.
The ppd is proportional to the product of a likelihood and a prior probability density, as

\begin{equation} \label{eq:Bayes}
\pi(\bm{\theta}|\bm{y}) \propto \pi(\bm{y}|\bm{\theta})\pi(\bm{\theta}),
\end{equation}

\noindent where $\bm{y}$ denotes the measurement data, and $\bm{\theta}$ the unknowns.
The likelihood $\pi(\bm{y}|\bm{\theta})$ includes the forward model that maps the parameters to the measurements, and information about the measurement noise and modelling uncertainty.
The prior probability density $\pi(\bm{\theta})$ is straightforward to construct based on physical constraints and information obtained from other sources such as previous experiments.

Once the posterior has been derived, we can compute parameter estimates such as the maximum a posteriori (MAP) or the conditional mean (CM) estimate

\begin{align}
\bm{\theta}_{\mathrm{MAP}} & = \arg\max_{\bm{\theta}}\pi(\bm{\theta}|\bm{y}), \\
\bm{\theta}_{\mathrm{CM}} & = \mathbb{E}\{\bm{\theta}|\bm{y}\} = \int\bm{\theta}\pi(\bm{\theta}|\bm{y})d\bm{\theta}. \label{eq:CMestimate}
\end{align}

\noindent We can also compute parameter uncertainty estimates, such as credible intervals.
A 95 \% credible interval $I_k(95) = [a_I,b_I]\subset\mathbb{R}$ for $\theta_k$ is defined as

\begin{equation}
\int_{a_I}^{b_I}\pi(\theta_k|\bm{y})d\theta_k = 0.95, \label{eq:95credIntval}
\end{equation}

\noindent where $\pi(\theta_k|y) = \int_{\mathbb{R}^{n_\theta-1}}\pi(\theta_1,\dots,\theta_{n_\theta}|\bm{y})d\theta_1\cdots\theta_{k-1}\theta_{k+1}\cdots\theta_{n_\theta}$ is the marginal density of the $k$-th component of $\bm{\theta}$, and $n_\theta$ is the number of unknowns in the model.
In the Bayesian framework, credible intervals can be directly interpreted as statements on the probabilities (i.e. uncertainties) of the parameter values.

\subsection{Likelihood}

The data vector $\bm{y}\in\mathbb{C}^{2n_\omega}$, where $n_\omega$ is the number of frequencies, consists of the measured reflection and transmission coefficients

\begin{equation}
\bm{y} = [\bm{R}, \bm{T}],
\end{equation}

\noindent where $\bm{R} = [R(\omega_1),\dots,R(\omega_{n_\omega})]$ and $\bm{T} = [T(\omega_1),\dots,T(\omega_{n_\omega})]$.
We assume that the data include complex valued additive measurement noise $\bm{e}$, which is circularly symmetric, i.e. its mean is zero and the real and imaginary parts are independent and have equal variance.
Then, with the usual assumption of mutual independence between $\bm{e}$ and $\bm{\theta}$, we can write the likelihood as

\begin{equation} \label{eq:likelihood}
\pi(\bm{y}|\bm{\theta}) \propto \det(\bm{\Gamma}_{\bm{e}})^{-C}\exp\left\{-C\norm{\bm{L}_{\bm{e}}(\bm{y} - f(\bm{\theta}))}^2\right\},
\end{equation}

\noindent where $f(\bm{\theta}):\mathbb{R}^{n_\theta}\rightarrow\mathbb{C}^{2n_\omega}$ is the forward model solved at frequencies $\omega_1,\dots,\omega_{n_\omega}$, $n_\theta$ is the number of unknown parameters, $\bm{\Gamma}_{\bm{e}}$ is the covariance matrix of the measurement noise, and $\bm{L}_{\bm{e}}$ is a matrix square root of the inverse of the noise covariance, i.e. $\bm{L}_{\bm{e}}^T\bm{L}_{\bm{e}}^{} = \bm{\Gamma}_{\bm{e}}^{-1}$.
Since our data are complex valued and circularly symmetric we have $C=1$, and Eq. \eqref{eq:likelihood} corresponds to the complex normal distribution \cite{lapidoth2017foundation}.
In the case of real valued data, we would have $C=1/2$.

We assume that the noise level of the measurements is unknown, but that it stays constant over the whole frequency range.
Further, the measurements of $\bm{R}$ and $\bm{T}$ are not expected to be correlated with each other because they are carried out separately, and dissipation in the medium prevents us from using conservation principles that could relate them.
The measurements may have differing noise levels and we can therefore write the diagonal measurement noise covariance matrix as

\begin{equation}
\bm{\Gamma}_{\bm{e}} = \begin{bmatrix}
\sigma_{e_R}^2 \bm{I} & 0 \\
0 & \sigma_{e_T}^2 \bm{I} \\
\end{bmatrix},
\end{equation}

\noindent where $\sigma_{e_R}^2$ and $\sigma_{e_T}^2$ denote the noise variance in the reflection and transmission measurements, respectively, and $\bm{I}$ is the $n_\omega\times n_\omega$ identity matrix.

\subsection{Errors related to the measurement}

The Bayesian approach makes it possible to take errors related to the measurement set-up into account.
These include, for example, errors in the positioning of the sample and/or transducers.
Kaczmarek \emph{et al.} \cite{kaczmarek2015ultrasonic} studied the errors of sample positioning in ultrasonic reflectometry, and found that errors in the position of the sample influence mainly the phase of $\bm{R}$, while errors in the sample inclination mainly affect the magnitude of $\bm{R}$.

In our case, to measure the phase of $\bm{R}$ exactly, we would need to make sure that the distance from the sound source to the ceramic plate is precisely the same as the distance from the source to the air-water interface which is used to record the reference signal.
However, there is always some uncertainty in measuring distances in an experimental setup.
Moreover, in a scanning system the distance to the object can change if the object is not perfectly straight or aligned along the scanning axis, or if the positioning system itself flexes slightly while moving.
We therefore take this possible discrepancy into account by multiplying the $\bm{R}$ given by the forward model with a distance correction term $\exp(-\I k_f\epsilon_R)$, where $\epsilon_R$ is the distance mismatch.
Multiplying by this distance term we can match the phase of the reflection coefficient to the measured one, and thus remove the dependence on the exact sample position.
As we will see in section~\ref{sec:results}, this parameter is identified independently of the other parameters and improves the accuracy of the model.

To account for an incorrect sample inclination (i.e. when the sample is not normal to the incoming ultrasound field), we would need to model the actual finite-sized transducers and the ultrasonic field they produce, instead of the current plane wave approximation.
Doing so would substantially increase the computational cost of the model, and render the current approach to the inverse problem infeasible.
One possibility to take the effects of any inclination error into account is to use the approximation error approach \cite{kaipio2007statistical,kaipio2013approximate}, but this was not pursued in the current paper.
We will discuss the validity of the normal incidence assumption in the current setup at the end of section~\ref{sec:results}.

Other parameters related to the measurements are the noise levels $\sigma_{e_R}$ and $\sigma_{e_T}$.
The noise level of a measurement is related to how accurately the unknown parameters can be estimated, and is therefore essential information.
We estimate the noise levels simultaneously with the other parameters.
Let us denote all the measurement uncertainty parameters by $\bm{\xi} = [L,\epsilon_R,\sigma_{e_R},\sigma_{e_T}]$.

\subsection{Prior density}
\label{ssec:prior}

\begin{table*}[t]
	\caption{Parameters of the prior density.}
	\label{tab:prior}
	\centering
	\begin{tabular}{lccccccccccc}
		\hline\hline
		\bfseries Parameter & \multicolumn{1}{c|}{$\Lambda$} &  \multicolumn{1}{c|}{$\alpha_\infty$} &  \multicolumn{1}{c|}{$\log_{10}k_0$} &  \multicolumn{1}{c|}{$\phi$} & \multicolumn{1}{c}{$K_{b,0}$, $N_0$} & \multicolumn{1}{c|}{$K_{s,0}$} & \multicolumn{1}{c|}{$Q_{K_b}$, $Q_{K_s}$, $Q_{N}$} & \multicolumn{1}{c|}{$\rho_s$} & \multicolumn{1}{c}{$L$} & \multicolumn{1}{c|}{$\epsilon_R$} & $\sigma_{e_R}$, $\sigma_{e_T}$ \\
		\bfseries Unit & \multicolumn{1}{c|}{$\mu$m} &  \multicolumn{1}{c|}{-} &  \multicolumn{1}{c|}{m$^2$} &  \multicolumn{1}{c|}{-} & \multicolumn{2}{c|}{GPa} & \multicolumn{1}{c|}{-} & \multicolumn{1}{c|}{kg$\cdot$m$^{-3}$} & \multicolumn{2}{c|}{mm} & - \\ \hline
		\bfseries Min & 0.001 & 1 & $-14$ & 0.01 & 0.01 & 0.01 & 1 & 1 & 10 & -3 & 0 \\
		\bfseries Mean & 30 &  2 & $-12$ & 0.4 & 15 & 30 & 50 & 2500 & 25 & 0 & 0.1 \\
		\bfseries Std & 20 &  1 & 2 & 0.1 & 10 & 20 & 50 & 1000 & 0.2 & 0.5 & 0.2 \\
		\bfseries Max & 1000 & 10 & $-8$ & 1 & 400 & 400 & 1000 & 6000 & 30 & 3 & 1 \\
		\hline\hline
	\end{tabular}
\end{table*}

Since the properties of the fluid ($K_f, \rho_f, \eta$) are assumed to be known, we have 11 unknown material parameters in the model.
Let us represent these by $\bm{\theta} = [\Lambda,\alpha_\infty,k_0,\phi,K_{b,0},Q_{K_b},K_{s,0},\\ Q_{K_s},N_0,Q_N, \rho_s]$.
To denote all the unknown parameters, we define $\tilde{\bm{\theta}} = [\bm{\theta}, \bm{\xi}]$.
We will use a prior that is a truncated normal distribution, i.e. $\tilde{\bm{\theta}} \sim \mathcal{N}(\tilde{\bm{\theta}}_*,\bm{\Gamma}_{\tilde{\bm{\theta}}}) \times B(\tilde{\bm{\theta}})$, where $\tilde{\bm{\theta}}_*$ denotes the mean, $\bm{\Gamma}_{\tilde{\bm{\theta}}}$ the covariance, and $B(\tilde{\bm{\theta}})$ is an indicator function:

\begin{equation} \label{eq:indicatorfcn}
B(\tilde{\bm{\theta}}) = 
\begin{cases}
1, & \text{if } \tilde{\theta_k} \in \text{physical bounds } \forall\: k,\\
0, & \text{otherwise.}
\end{cases}
\end{equation}

\noindent The indicator function returns a probability one if all parameters are within the predefined bounds, and otherwise a probability of zero.
The bounds are based on considering what values are physically possible, and are found in Table~\ref{tab:prior}.

The prior mean is chosen based on available information from the manufacturer \cite{filtrosManual}, other studies \cite{johnson1994probing}, and our experience with similar materials.
For example, the manufacturer reports typical data on the porosity, pore size, and permeability of QF-20.
However, the prior knowledge is still limited, and we adjust the prior variance so that all expected parameter values have some probability.
We do not assume correlations between the parameters a priori, so the matrix $\bm{\Gamma}_{\tilde{\bm{\theta}}}$ is diagonal.
The prior mean and variance are also given in Table~\ref{tab:prior}.
Permeability is expressed on a logarithmic scale since it can take values that span multiple orders of magnitude \cite{schon2015physical}.

In addition to minimum and maximum values, the indicator function can also be used to impose other prior constraints.
For example, from the physical point of view, we know that the frame of a porous material is either rigid (it does not move under pressure), limp (it does not resist movement at all), or something in between.
This condition can be expressed as $0\le K_{b,0}\le K_{s,0}(1 - \phi)$ \cite{biot1957elastic}, which restricts the possible values of $K_{b,0}$ and $K_{s,0}$.
Other physical constraints we impose are that the Poisson's ratio of the porous frame is non-negative, which can be expressed as $K_{b,0} \ge 2/3N_0$, and that attenuation in the frame is greater than attenuation in pure solid, $Q_{K_b}^{-1} > Q_{K_s}^{-1}$.

\subsection{Sampling the posterior using MCMC}

Considering the form of the likelihood \eqref{eq:likelihood} and the prior, the posterior \eqref{eq:Bayes} can now be written out.
For numerical reasons, it is preferable to work with the logarithm of the posterior, which reads

\begin{equation} \label{eq:log_posterior}
\begin{split}
\log\pi(\tilde{\bm{\theta}}|\bm{y}) \propto &-\left\Vert \bm{L}_e(\bm{y} - f(\tilde{\bm{\theta}}))\right\Vert^2 -2n_{\omega}(\log\sigma_{e_R} + \log\sigma_{e_T}) \\
& - \frac{1}{2}\Vert \bm{L}_{\tilde{\bm{\theta}}}(\tilde{\bm{\theta}} - \tilde{\bm{\theta}}_*)\Vert^2 + \log B(\tilde{\bm{\theta}}),
\end{split}
\end{equation}

\noindent where we have denoted $\bm{L}_{\tilde{\bm{\theta}}}^{}\bm{L}_{\tilde{\bm{\theta}}}^T = \bm{\Gamma}_{\tilde{\bm{\theta}}}^{-1}$.

In this work, we use the CM, Eq. \eqref{eq:CMestimate}, as the parameter point estimate, and the 95 \% credible interval, Eq. \eqref{eq:95credIntval}, as the uncertainty estimate.
Computing these estimates requires the solving of high-dimensional integrals, which are in practice often approximated with Markov chain Monte Carlo (MCMC) \cite{metropolis1953equation} methods.
MCMC methods generate an ensemble of samples that are distributed according to the ppd, and given these samples it is easy to compute point and interval estimates for the parameters.
For a general reference on MCMC methods, see e.g. Ref.~\cite{brooks2011handbook}.

MCMC methods are computationally heavy, and the efficiency of the sampler plays a big part on whether sampling methods are viable in a given problem.
Here we use the sampler developed in \cite{niskanen2019characterising}, which is based on an adaptive random walk Metropolis algorithm \cite{haario2001adaptive,andrieu2008tutorial}, with parallel tempering \cite{geyer1991markov,earl2005parallel}.
We use 10 temperatures, two of which are set to one and the rest are adapted during the MCMC run to optimise efficiency.
The first 15,000 samples are removed as burn-in, and the sampler is run until an objective convergence criterion is fulfilled.
Our stopping criterion is based on computing the Monte Carlo Standard Error \cite{flegal2008markov} for each parameter and ensuring that it is small enough compared to the posterior variance of the parameter.
On average, the stopping criterion was reached in 35,000 samples after burn-in, and the multivariate effective sample size \cite{vats2019multivariate} was 850.

\section{Characterisation results}
\label{sec:results}

\begin{figure*}[t]
	\centering
	\includegraphics[width=0.4\linewidth]{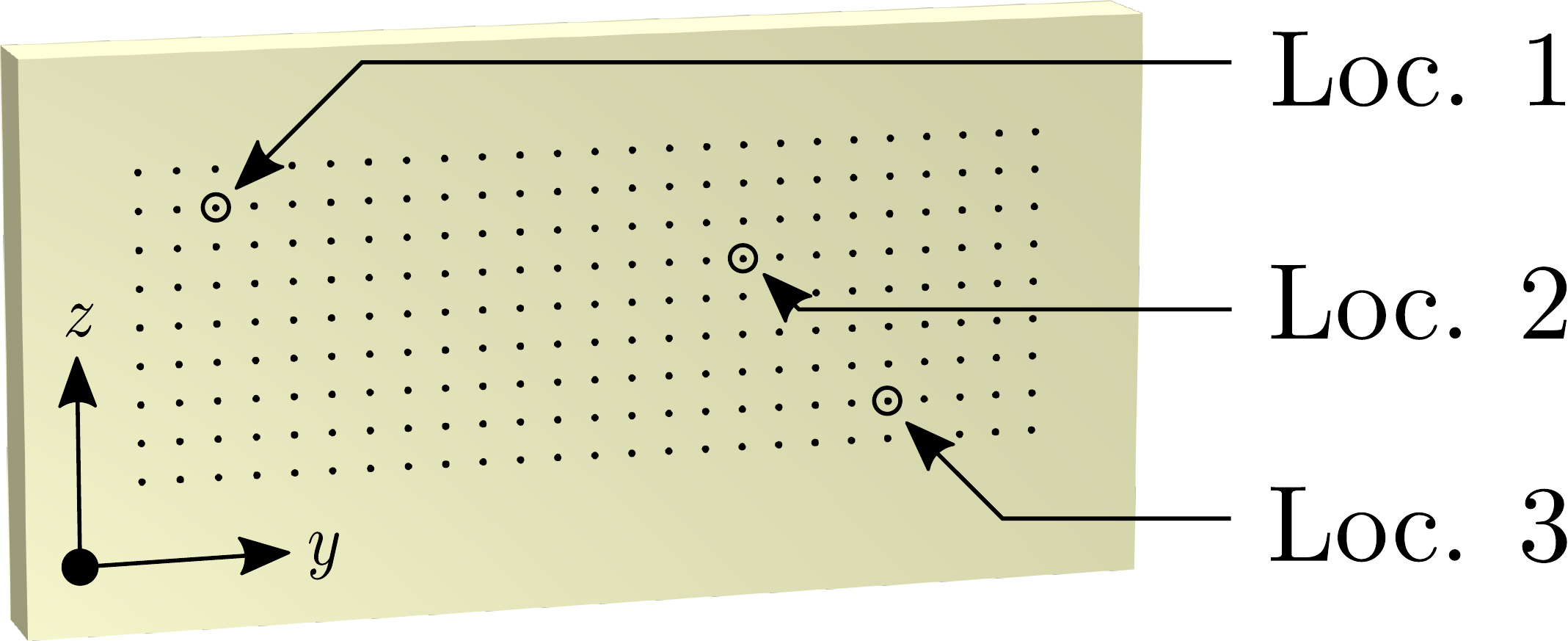}
	\caption{Three locations selected for a closer analysis.}
	\label{fig:example_locations}
\end{figure*}

Let us now present and discuss the results of the inversion.
Due to the large number of measurement locations, we first focus on a few selected ones, shown in Fig.~\ref{fig:example_locations}, and then summarise all results by plotting the CM and uncertainty estimates of each parameter as two dimensional maps.

\subsection{Individual locations}

\begin{figure*}[th!]
	\centering
	\includegraphics[width=\linewidth]{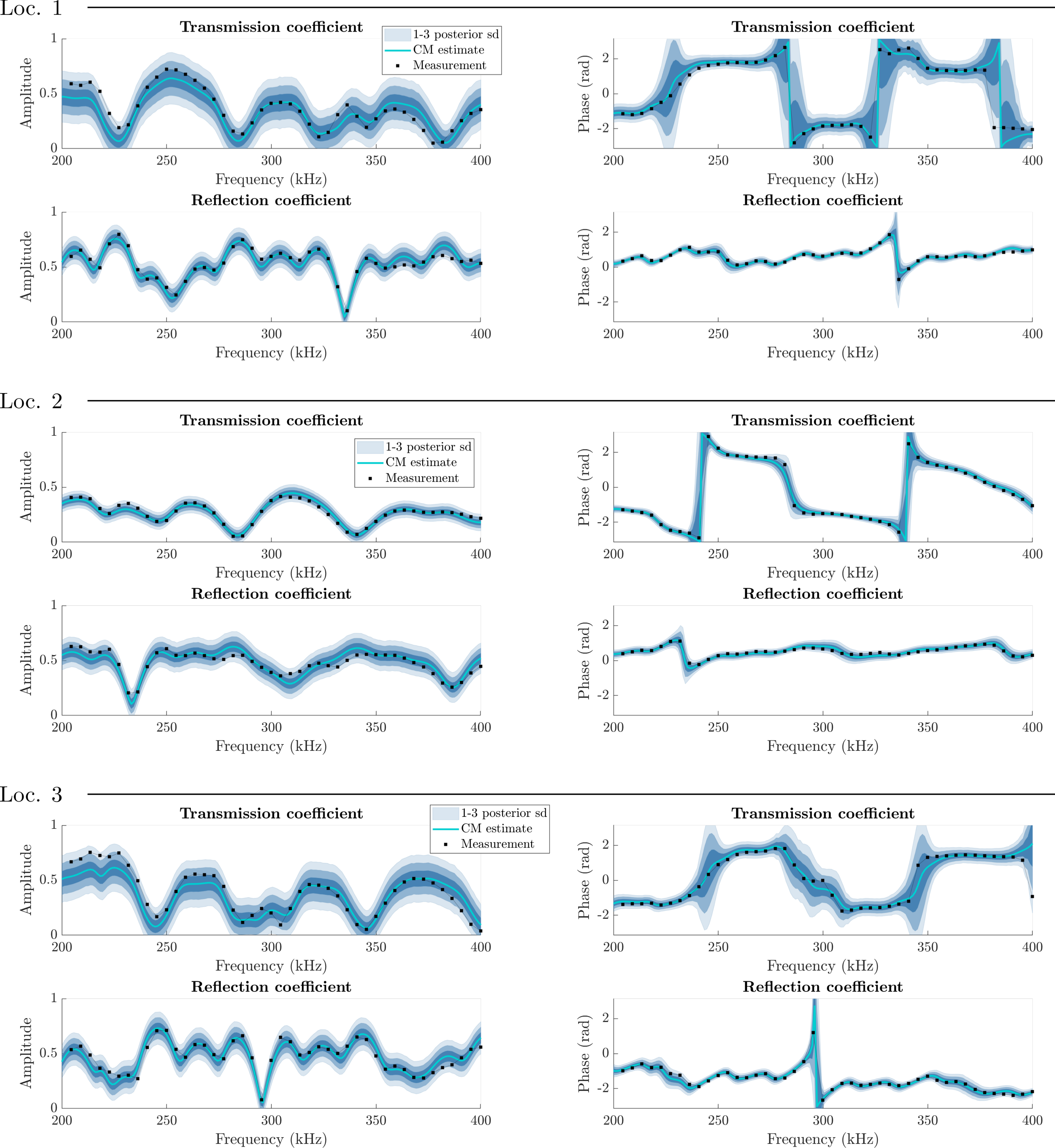}
	\caption{Measurement, model fit corresponding to the CM estimate, and posterior predictive distribution as standard deviations, at the three example locations. Range of the phase plots is limited to $[-\pi,\pi]$, but the standard deviations may continue outside the window.}
	\label{fig:all_modelfits}
\end{figure*}

The measured reflection and transmission coefficients vary significantly depending on where on the ceramic plate the measurement is taken.
This already shows that the plate is likely not homogeneous in the macroscopic sense, and that, if accurate knowledge on the parameters is needed, one should avoid averaging measurements over several locations.
Fig.~\ref{fig:all_modelfits} demonstrates the variability in the measurements by showing the magnitude and phase of the measured $\bm{R}$ and $\bm{T}$ coefficients from three locations around the object.
In addition, the figure shows the model prediction corresponding to the CM estimate, and indicates the range of the model predictions with shading that corresponds to one, two, and three standard deviations.
Variance of the model predictions is mainly induced by the noise, which by our assumption is normally distributed.
Therefore standard deviations describe the distribution of the model predictions well.

As Fig.~\ref{fig:all_modelfits} shows, the frequency regions of almost zero reflection or transmission are different in all three locations, and the goodness of the model fit differs from location to location.
In locations 1 and 3, the model underestimates the transmission coefficient amplitude at low frequencies, whereas in location 2 the model fit to the transmission coefficient is excellent over the whole frequency range.
The goodness of the model fit can be evaluated from the width of the model prediction standard deviations.
What should be noted, however, is how well the Biot model fits to the measurements, despite the variability in the measurement data and simplifications (such as the plane wave assumption) in the forward model.

\begin{table*}[ht]
	\caption{Conditional mean (\textbf{CM}, bold for easier readability) and 95 \% credible interval [CI$_a$ -- CI$_b$] estimates for the example locations (see Fig.~\ref{fig:example_locations}).}
	\label{tab:inversion_results}
	\centering
	\begin{tabular}{c|ccc|ccc|ccc|cc}
		\hline \hline
		\multicolumn{1}{c}{\bfseries Parameter} & \multicolumn{3}{c}{\shortstack{Loc. 1\\CI$_a$ $\:$  \textbf{CM} $\:$ CI$_b$}} & \multicolumn{3}{c}{\shortstack{Loc. 2\\CI$_a$ $\:$  \textbf{CM} $\:$ CI$_b$}} & \multicolumn{3}{c}{\shortstack{Loc. 3\\CI$_a$ $\:$  \textbf{CM} $\:$ CI$_b$}} & \bfseries Ref. \cite{johnson1994probing}\footnotemark[1] & \bfseries Ref. \cite{filtrosManual}\footnotemark[2] \\
		\hline
		$\Lambda$ ($\mu$m)     & 47.7  & \textbf{66.9} & 90.2  & 36.8  & \textbf{53.7}  & 70.1  & 64.7  & \textbf{89.2}  & 113.0 & 19.0 & 84\footnotemark[3] \\
		$\alpha_\infty$        & 2.13  &\textbf{2.20}  & 2.25  & 2.05  & \textbf{2.13}  & 2.17  & 2.02  & \textbf{2.10}  & 2.16  & 1.89 & \\
		$\log_{10}k_0$ (m$^2$) & -11.3 &\textbf{-10.2} & -8.5  & -11.4 & \textbf{-10.2} & -8.6  & -11.0 & \textbf{-9.9}  & -8.4  & -7.77 & -10.7 -- -10.4 \\
		$\phi$                 & 0.33  &\textbf{0.34}  & 0.36  & 0.32  & \textbf{0.34}  & 0.35  & 0.36  & \textbf{0.38}  & 0.40  & 0.402 & 0.35 -- 0.45 \\
		$K_{b,0}$ (GPa)        & 12.0  &\textbf{13.2}  & 14.5  & 9.8   & \textbf{10.5}  & 11.4  & 9.3   & \textbf{10.6}  & 12.0  & 9.47 & \\
		$Q_{K_b}$              & 11.3  &\textbf{14.3}  & 18.0  & 6.7   & \textbf{7.7}   & 8.8   & 10.1  & \textbf{12.5}  & 15.3  & & \\
		$K_{s,0}$ (GPa)        & 18.7  &\textbf{21.8}  & 26.5  & 15.3  & \textbf{17.0}  & 19.2  & 15.7  & \textbf{19.3}  & 24.0  & 36.6 & \\
		$Q_{K_s}$              & 45.3  &\textbf{105.1} & 171.8 & 28.2  & \textbf{88.8}  & 155.1 & 21.3  & \textbf{75.8}  & 142.1 & & \\
		$N_0$     (GPa)        & 17.3  &\textbf{18.7}  & 20.1  & 14.1  & \textbf{15.0}  & 15.9  & 12.7  & \textbf{14.4}  & 15.9  & 7.63 & \\
		$Q_N$                  & 71.3  &\textbf{134.4} & 203.8 & 58.5  & \textbf{117.6} & 177.8 & 58.7  & \textbf{115.1} & 177.3 & & \\
		$\rho_s$ (kg$\cdot$m$^{-3}$) & 2824  &\textbf{3053}  & 3296  & 2613  & \textbf{2733}  & 2855  & 2880  & \textbf{3133}  & 3398  & 2760 & \\
		\hline\hline
	\end{tabular}
	\footnotetext[1]{Presented as an order of magnitude illustration only.}
	\footnotetext[2]{Typical values stated by the manufacturer.}
	\footnotetext[3]{Nominal particle size $a = 20$ $\mu$m. $\Lambda = 8k_0\alpha_\infty/(a\phi)$, with $k_0 = 10^{-10.4}$ m$^2$, $\alpha_\infty = 2.1$, and $\phi = 0.4$. See Eq. (28) in Ref. \cite{niskanen2019characterising}.}
\end{table*}

Let us now consider to what degree does the variability in the measurements show as variability of the parameter estimates.
Although porous materials are inherently heterogeneous, it is expected that measurements of the same material should lead to similar parameter estimates.
The computed CM and 95 \% CI estimates for locations 1--3 are shown in Table~\ref{tab:inversion_results}.
The table also shows data from the manufacturer \cite{filtrosManual}, and some results of other characterisation processes and some textbook values reported by Johnson \emph{et al.} \cite{johnson1994probing}, shown here only as an illustration of the order of magnitude some parameters may take.
These are not expected to coincide exactly to the currently estimated parameters, due to heterogeneities in the material, different frequency ranges, and uncertainties associated with the direct characterisation methods.
In addition, Johnson \emph{et al.} do not consider attenuation in the solid.
We can see from Table~\ref{tab:inversion_results} that the estimated parameter values do vary with location, but in most cases the variability is within the 95 \% uncertainty bounds.

\begin{figure*}[t]
	\centering
	\includegraphics[width=\linewidth]{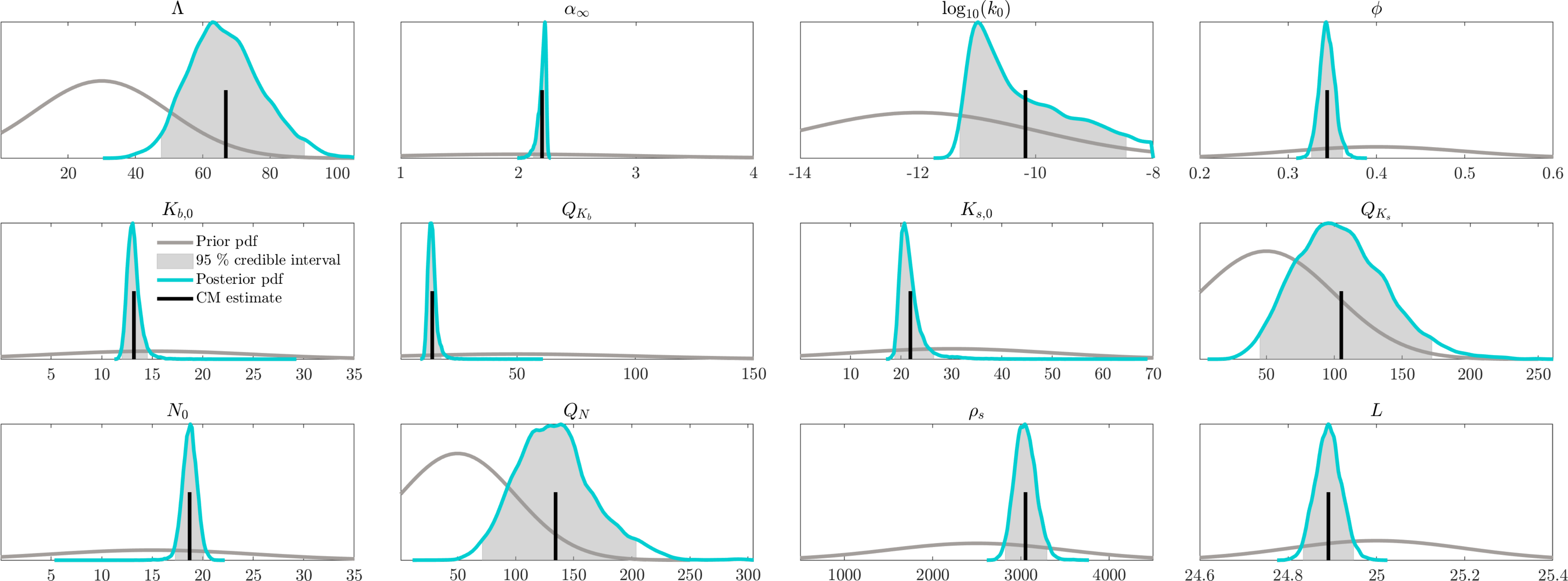}
	\caption{Posterior marginal distributions, measurement Loc.~1.}
	\label{fig:loc1_1D}
\end{figure*}

Sampling the ppd using MCMC gives us access to the marginal and joint marginal posterior densities, which are helpful in assessing the identifiability of each model parameter.
Identifiability, however, is not a straightforward concept to define, and depends on both the model and data.
One way to think of identifiability is in terms of posterior variance, and how much does incorporating the data reduce the posterior variance compared to the prior variance.
A visual representation of posterior and prior variances is found in Fig.~\ref{fig:loc1_1D}, where the marginal posterior distributions of measurement location 1 are drawn over the marginal prior distributions.
According to the criteria of posterior variance reduction, the parameters $\alpha_\infty, \phi, K_{b,0}, Q_{K_b}, K_{s,0}, N_0$, and $\rho_s$ are the best identified, while the posterior of the rest of the model parameters is more similar to the prior.
However, no parameters have a marginal posterior exactly like the prior, which shows that the data carry some information on all parameters.
The reasonably low uncertainties of many parameters show that the measurement data, which include contributions from both the fast and the slow wave, contain a lot of information.

\begin{figure*}[t]
	\centering
	\includegraphics[width=\linewidth]{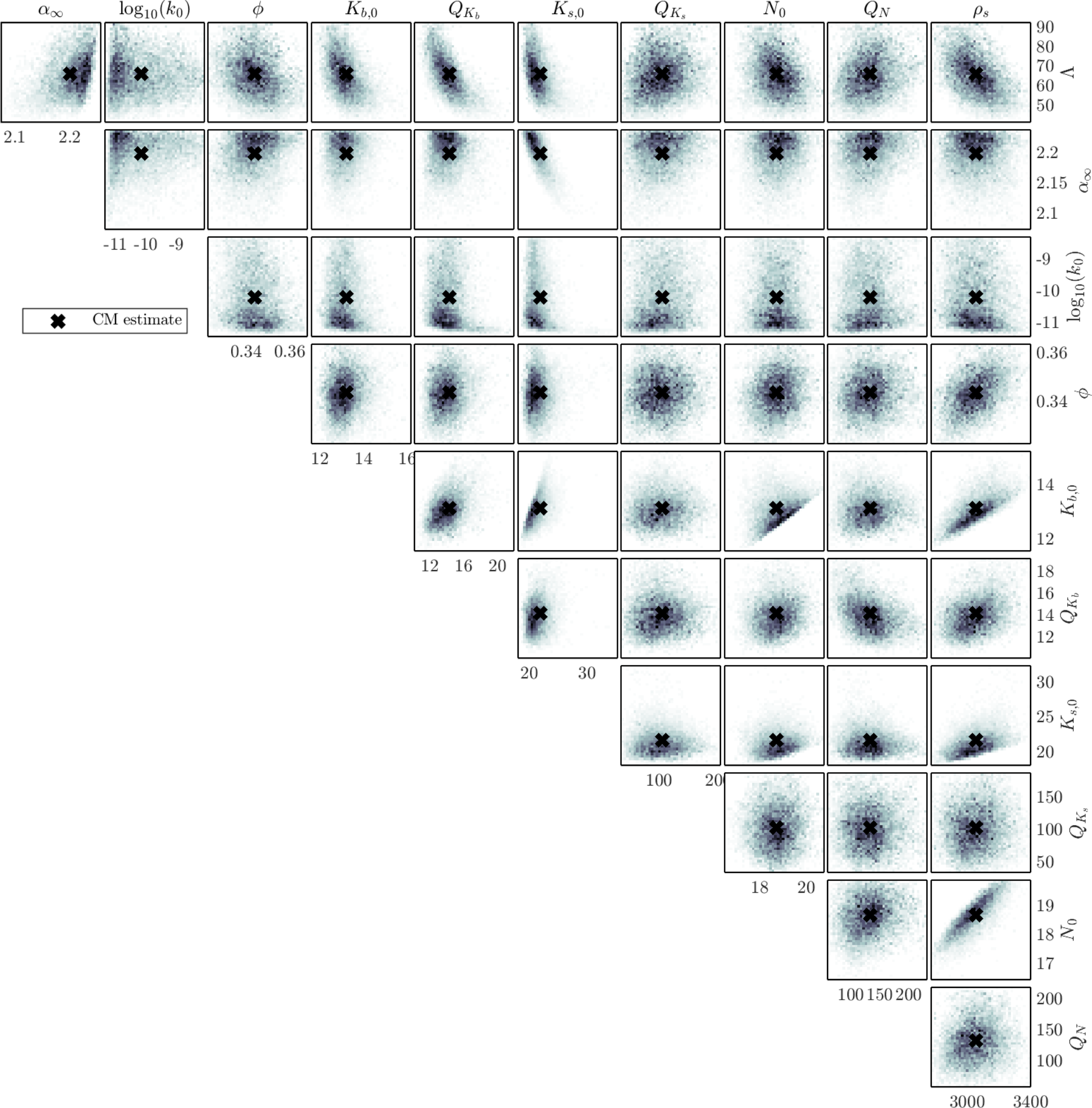}
	\caption{Joint posterior marginal distributions, measurement Loc.~1.}
	\label{fig:loc1_2D}
\end{figure*}

Inspecting the joint densities can reveal other type of identifiability issues, namely strong correlations between parameters.
Fig.~\ref{fig:loc1_2D} shows an example of the joint marginal posterior distributions of the model parameters.
Again, the results are from measurement location 1, but the findings are representative of the whole measurement set.
Because all the measurements are at normal incidence where the shear wave is not generated, we can expect that the frame bulk modulus $K_{b,0}$ and the shear modulus $N_0$ are correlated and not necessarily identifiable separately \cite{niskanen2019characterising}.
Fig.~\ref{fig:loc1_2D} shows that these parameters do exhibit some correlation, but there is also a sharp cut in their joint marginal posterior.
This cut is the result of adding to the prior the requirement that the Poisson's ratio has to be positive, and the line can be seen to occur at $K_{b,0} = 2/3N_0$.
The strongest correlations are between $\rho_s$ and $N_0$, as well as between $\rho_s$ and $K_{b,0}$.
Also the solid bulk modulus $K_{s,0}$ exhibits correlation with several parameters.

\subsection{Spatial variability of the parameter estimates}

\begin{figure*}[t]
	\centering
	\includegraphics[width=\linewidth]{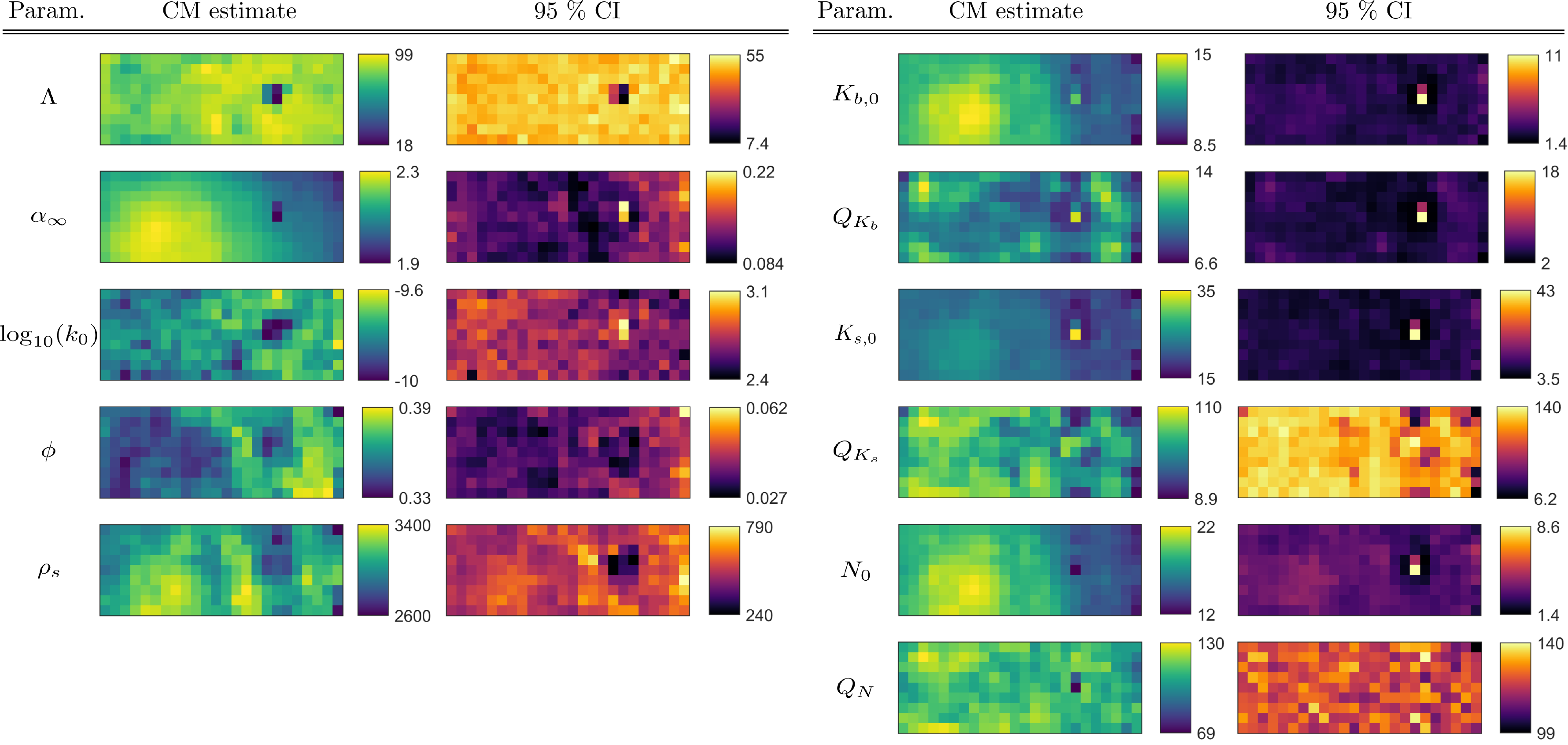}
	\caption{CM estimates of the Biot model parameters, and the widths of the related 95 \% credible intervals, at 216 measurement locations. Each pixel of a parameter map corresponds to a measurement location on the plate (see Fig.~\ref{fig:example_locations}), and the units are the same as in Table~\ref{tab:prior}.}
	\label{fig:map_CMandCI}
\end{figure*}

Let us now examine the results from all the measured locations together.
To reveal possible spatial structures or variations in the measured plate, we plot the inversion results as a two dimensional map.
The first and third columns of Fig.~\ref{fig:map_CMandCI} show the CM model parameter estimates, and the second and fourth columns show the uncertainty related to each CM estimate as the width of the 95 \% CI.
Each subplot consists of a $9\times24$ grid of pixels, and the value of each pixel corresponds either to the CM or CI computed based on the measurement made in that location.

Figure~\ref{fig:map_CMandCI} shows that the parameter estimates vary smoothly from point to point.
Because the inversion is carried out individually at each point, with no connection to the neighbouring measurements, we can conclude that the smoothness is found in the measurement data.
Furthermore, if the spatial changes of all parameter estimates were similar in smoothness to the one of $\alpha_\infty$, for example, we could argue that perhaps the effective measurement area of one transducer is much larger than the -6 dB diameter of 10 mm, and the smoothness is the result of low measurement resolution.
The scale of changes in $\alpha_\infty$ would point to a effective measurement diameter of about 10 cm.
However, parameters such as $\phi$ and $\rho_s$ vary smoothly as well, but in a much smaller scale.
This points toward a conclusion that we do have a resolution of 10--20 mm, in line with the geometrical analysis in section~\ref{sec:experiments}, and that the large smooth areas of some parameters are a property of the object.
In conclusion, results in Fig.~\ref{fig:map_CMandCI} show that the measured object is inhomogeneous, and that there is a gradual change in the porous frame going from left to right.
When bulk and shear moduli decrease, porosity increases, which is consistent with the effective medium theory of porous ceramics \cite{munro2001effective}.

Another notable feature of Fig.~\ref{fig:map_CMandCI} is a small spot in the middle-right side of the plate, where $\Lambda$ and $\alpha_\infty$ are clearly smaller than in the other locations, whereas $K_{s,0}$ is much larger than the average.
In addition, the estimated parameter uncertainty at that location is several times higher than elsewhere for many parameters.
This result points to an anomaly in the plate, and it is interesting that it can be seen in basically every parameter.
We selected this anomaly as the example location 2, and the model fit and parameter estimates at this point were considered earlier.

\subsection{Uncertainty parameters}

An integral part in carrying out the parameter estimation in the Bayesian framework is the ability to also account for the so called nuisance parameters, parameters that we are not interested in but nevertheless affect the result of the inversion.

\begin{figure}[t]
	\centering
	\includegraphics[width=0.5\linewidth]{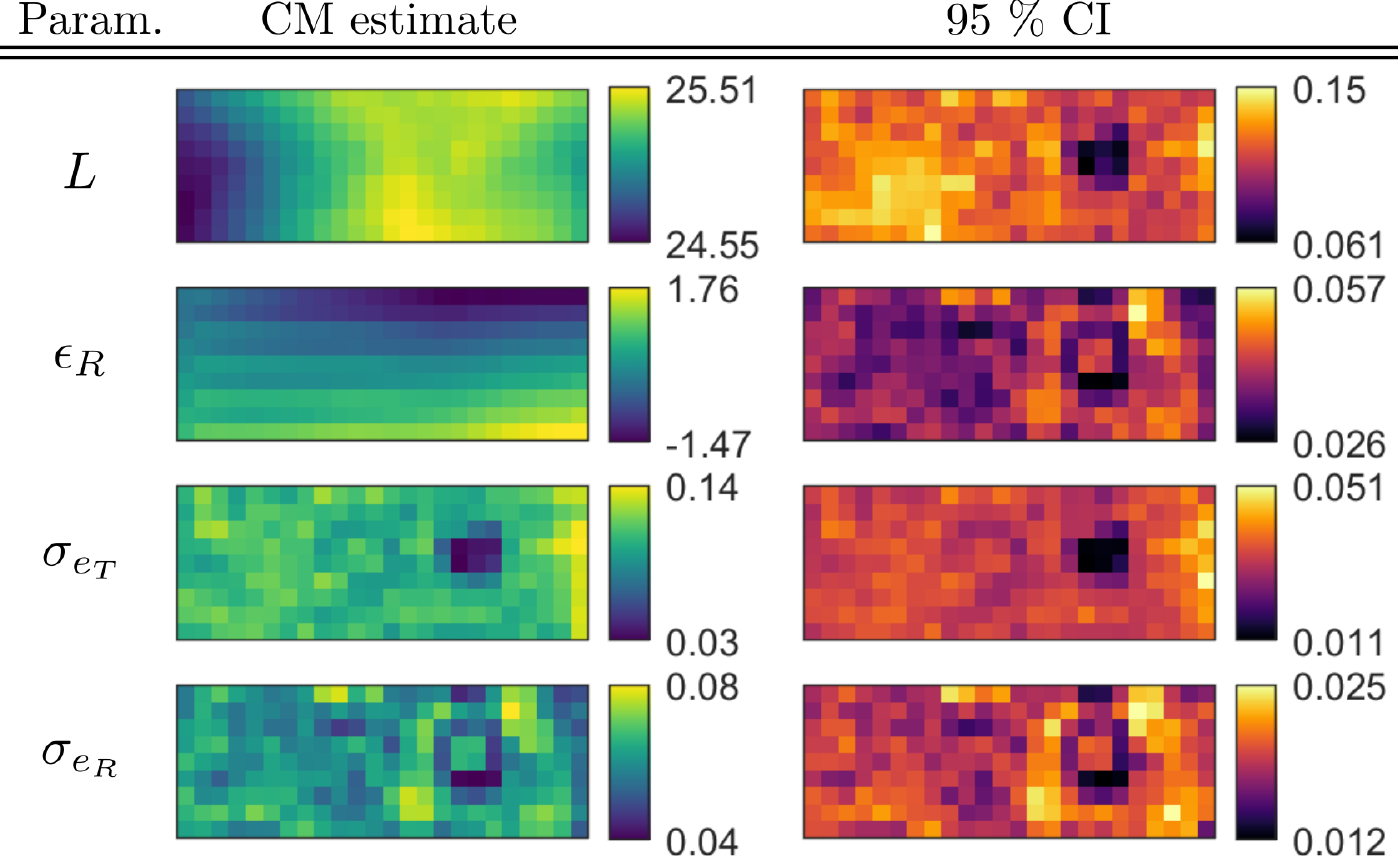}
	\caption{CM estimates of the measurement uncertainty parameters, and the widths of the related 95 \% credible intervals. Units are the same as in Table~\ref{tab:prior}.}
	\label{fig:map_CMandCI_nuisance}
\end{figure}

Fig.~\ref{fig:map_CMandCI_nuisance} shows the CM and CI estimates of the measurement uncertainty parameters $\bm{\xi}$ considered in this paper.
Let us first comment on the estimated measurement noise levels.
These parameters include not only the random white noise component but also any model errors, which is why the estimated noise levels vary around the plate.
For example, in location 2 the estimated transmission noise level is lower than at any other point, which is confirmed by the good model fit seen in Fig.~\ref{fig:all_modelfits}.
Apart from the small area around location 2, the noise levels do not change much.
This shows that the Biot model fits equally well to measurements from all over the object.

We can also see that the measured plate does not have perfectly parallel faces, but its thickness varies up to almost a millimetre along the plate.
The associated uncertainty interval is 0.1 mm on average, which shows that the data provide accurate information on the thickness.
The uncertainties in the estimates of $L$ follow a pattern similar to the estimates of the transmission measurement noise level $\sigma_{e_T}$, which suggests that most of the information on thickness is found in the transmission coefficient.
This can be explained by the fact that the phase of $\bm{T}$ is directly linked to $L$, as can be seen from \eqref{eq:Wiener_filter_T}, whereas the link between $\bm{R}$ and $L$ is more implicit and thickness can be correlated with other parameters.
We can also rule out changes in thickness as the reason for the smooth change in some parameter estimates, since the way the thickness estimates change is different to the change in the Biot parameters.
The varying thickness is another reason why we could not use synthetic plane wave techniques in the measurements.
It also shows that using a constant value of thickness in the measurements would add modelling error.

The estimates of $\epsilon_R$ are also very accurate, since the maximum 95 \% CI for the estimates is less than 0.06 mm.
However, the distance from the reflection transducer to the plate changes by over a millimetre, and without accounting for this distance mismatch the model fit would not be nearly as good.
Interestingly, we can even see the scanning pattern the measurement device has taken, and that when scanning from left to right the transducer has been slightly closer to the plate than when scanning in the opposite direction.

The estimated uncertainty parameters give detailed information on the inclination of the plate which we can use to assess the validity of the normal incidence assumption.
First, changes in $\epsilon_R$ tell us about the orientation of the plate's closest face with respect to the reference plane the measurement system moves in.
Fig.~\ref{fig:map_CMandCI_nuisance} shows that if we start from the centre of the plate, moving horizontally the plate is approximately parallel to the reference plane, whereas vertically the plate is at a small angle, with the top leaning away.
This can be compensated by tilting the transducers towards the normal of the plate.
However, the plate tilt angle changes from the left hand side to the right, and some error is inevitably introduced.
Based on the estimates of $\epsilon_R$, the difference in the vertical tilt angle of the plate between the left and right sides is $1.5$ degrees.
We can also calculate the tilt angle that the changes in $L$ produce.
As a conservative estimate, assuming that one side of the plate is flat, a 1 mm change in thickness over the distance from the thinnest to the thickest part corresponds to about $0.5$ degree tilt from the reference plane.

Numerical simulations show that the relative difference between measurements at normal and oblique incidences are less than $2$ \% when the deviation is one degree, but typically between $3$ and $6$ \% when the deviation is two degrees.
Thus the assumption of normal incidence may induce some error into the inversion, but this error is small compared to the deviations between the actual measurements and the best model prediction (as in Fig.~\ref{fig:all_modelfits}).
Possible reasons for the observed of model discrepancy include depth-wise heterogeneities in the plate, and the plane wave approximation, where we are modelling the focused ultrasound field as a single plane wave.
Investigating the possible bias this approximation introduces to the parameter estimates would be an interesting topic of future research.

\section{Conclusion}
\label{sec:conclusion}

In this paper, we estimated the physical parameters of a poroelastic (Biot) object using only ultrasonic reflection and transmission measurements made in a water tank.
We measured over 200 different points on the object to assess how the parameters change spatially.
The inverse problem was solved in the Bayesian framework, which allowed us to account for measurement and model errors, and to quantify the uncertainty related to the parameter estimates.
The parameter inference was carried out using a Markov chain Monte Carlo algorithm.
We found that the computational model described the measurements well, and that the measured data carried information on every parameter in the Biot model.
With the proposed method, we were able to identify spatial changes in the parameters along the object, and provide uncertainty estimates for all parameters.

\section*{Acknowledgement}
This work has been supported by the strategic funding of the University of Eastern Finland, by the Academy of Finland (Finnish Centre of Excellence of Inverse Modelling and Imaging, and project 321761), by the RFI Le Mans Acoustique (Pays de la Loire) Decimap project, and the Jenny and Antti Wihuri Foundation. This article is based upon work initiated under the support from COST Action DENORMS CA-15125, funded by COST (European Cooperation in Science and Technology).

\bibliographystyle{unsrt}

\end{document}